\documentclass[%
 reprint,
 amsmath,amssymb,
 aps,
 prl,
superscriptaddress,
showpacs,
]{revtex4-1}

\usepackage{graphicx}
\usepackage{dcolumn}
\usepackage{bm}

\begin{document}

\preprint{APS/123-PRL}

\title{Normal-dispersion Microcombs Enabled by Controllable Mode Interactions}

\author{Xiaoxiao Xue}
 \email{xiaoxxue@gmail.com}
 \affiliation{School of Electrical and Computer Engineering, Purdue University, 465 Northwestern Avenue, West Lafayette,
Indiana 47907-2035, USA.}

\author{Yi Xuan}%
 \affiliation{School of Electrical and Computer Engineering, Purdue University, 465 Northwestern Avenue, West Lafayette,
Indiana 47907-2035, USA.}
 \affiliation{Birck Nanotechnology Center, Purdue University, 1205 West State Street, West Lafayette, Indiana 47907, USA.}

\author{Pei-Hsun Wang}
 \affiliation{School of Electrical and Computer Engineering, Purdue University, 465 Northwestern Avenue, West Lafayette,
Indiana 47907-2035, USA.}

\author{Yang Liu}
 \affiliation{School of Electrical and Computer Engineering, Purdue University, 465 Northwestern Avenue, West Lafayette,
Indiana 47907-2035, USA.}

\author{Dan E. Leaird}%
 \affiliation{School of Electrical and Computer Engineering, Purdue University, 465 Northwestern Avenue, West Lafayette,
Indiana 47907-2035, USA.}

\author{Minghhao Qi}%
 \affiliation{School of Electrical and Computer Engineering, Purdue University, 465 Northwestern Avenue, West Lafayette,
Indiana 47907-2035, USA.}
 \affiliation{Birck Nanotechnology Center, Purdue University, 1205 West State Street, West Lafayette, Indiana 47907, USA.}

\author{Andrew M. Weiner}%
 \email{amw@purdue.edu}
 \affiliation{School of Electrical and Computer Engineering, Purdue University, 465 Northwestern Avenue, West Lafayette,
Indiana 47907-2035, USA.}
 \affiliation{Birck Nanotechnology Center, Purdue University, 1205 West State Street, West Lafayette, Indiana 47907, USA.}


\begin{abstract}
We demonstrate a scheme incorporating dual coupled microresonators through which mode interactions are intentionally introduced and controlled for Kerr frequency comb (microcomb) generation in the normal dispersion region. Microcomb generation, repetition rate selection, and mode locking are achieved with coupled silicon nitride microrings controlled via an on-chip microheater. Our results show for the first time that mode interactions can be programmably tuned to facilitate broadband normal-dispersion microcombs. The proposed scheme increases freedom in microresonator design and may make it possible to generate microcombs in an extended wavelength range (e.g., in the visible) where normal material dispersion is likely to dominate.


\end{abstract}

\maketitle

Microresonator-based Kerr frequency comb (microcomb) generation is a promising technique with many advantages including very compact size, high repetition rate, and (sometimes) wide bandwidth \cite{r6}. Many applications can potentially benefit from this revolutionary technique, such as optical communications \cite{r3}, photonic radio-frequency signal processing \cite{r4}, and optical clocks \cite{r5}.
The initial stage of comb formation inside the microresonator is triggered by modulational instability which is enabled by Kerr nonlinearity and group velocity dispersion. Anomalous dispersion has been widely exploited \cite{r1,r7,r8,r9,r10,r11,r12,r13,r14,r15,r16,r17,r18}, while normal dispersion has often been thought unsuitable for microcomb generation due to the lack of modulational instability on the upper branch of the bistable pump field in the normal dispersion region \cite{r19,r20,r21}. Nevertheless, comb generation in normal-dispersion microresonators has by now been reported several times \cite{r22,r23,r24,r25,r26,r27,r28,r29,r30}. Normal-dispersion microcombs are of particular interest as normal dispersion is easy to access in most nonlinear materials. It has been found that modulational instability may occur in the normal dispersion region with the aid of coupling between different family modes \cite{r24,r28,r29,r34}. However, the mode interactions employed so far in a single microresonator are related to accidental degeneracies associated with different spatial modes and are hence very difficult to control and modify. In this letter, we demonstrate a novel scheme which for the first time permits programmable and reliable control of mode interactions to achieve repetition-rate-selectable and mode-locked combs from normal dispersion microresonators.

\begin{figure}[btp]
\includegraphics[width=8.3cm]{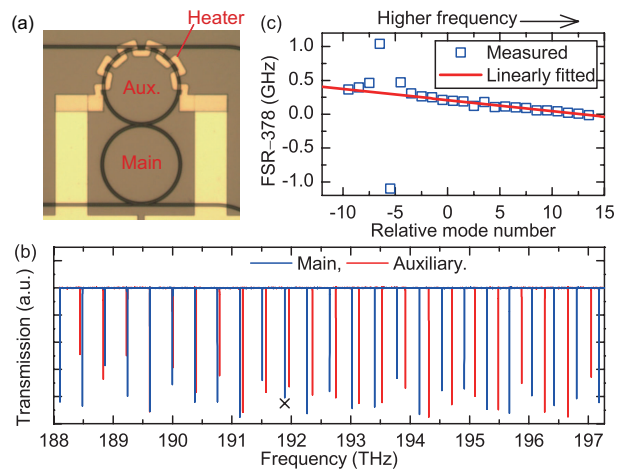}
\caption{\label{fig:device} Dual coupled silicon nitride microrings. (a) Microscope image. The radii of the main and auxiliary rings are $60\ \mathrm{\mu m}$ and $58\ \mathrm{\mu m}$, respectively. An integrated microheater shifts the resonances of the auxiliary ring via the thermo-optic effect. (b) Linear transmission spectra measured with a low-power frequency scanning laser at the bus waveguide of either the main or auxiliary ring. The FSRs of the main and auxiliary rings are $378\ \mathrm{GHz}$ and $391\ \mathrm{GHz}$, respectively. (c) FSR versus the relative mode number for the main ring. Mode 0 corresponds to the resonance around 191.9 THz which is marked in subplot (b) with $\times$. The dispersion is $\beta_2 \approx 130\ \mathrm{ps^2/km}$, which is extracted from the slope of the FSR curve in subplot (c). The jumps in FSR around mode $-6$ are due to the coupling between the two rings in the mode crossing region (around 189.5 THz in subplot (b)).}
\end{figure}

\begin{figure*}[btp]
\includegraphics[width=16.5cm]{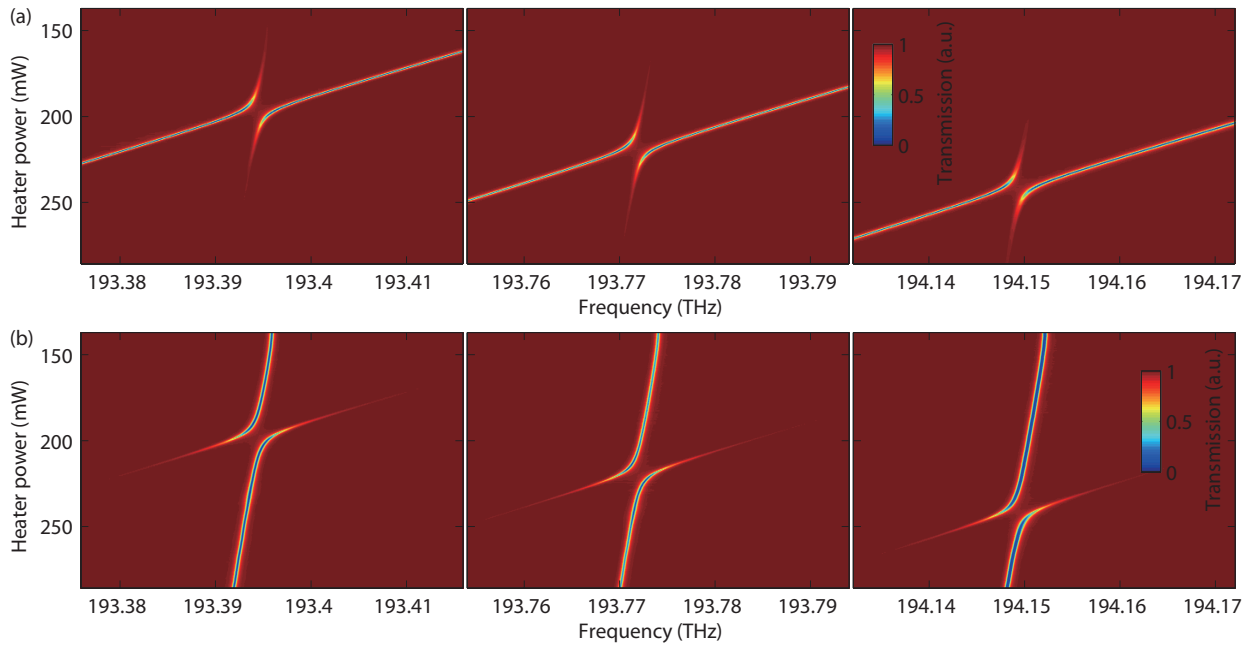}
\caption{\label{fig:trans_2D} Transmission spectra versus heater power, showing that the resonances can be selectively split. (a) Auxiliary ring; (b) main ring. Three consecutive resonances are shown. The thermal tuning efficiency of the auxiliary ring is $-600\ \mathrm{MHz/mW}$, while the main ring shifts only slightly at a rate of $-30\ \mathrm{MHz/mW}$. By changing the heater power, the strongest mode interaction moves from one resonance to the next.}
\end{figure*}

Figure \ref{fig:device}(a) shows the microscope image of our device which is based on the well-known coupled-microresonator structure \cite{r31,r32}. Two silicon nitride (SiN) microrings (main and auxiliary) are coupled to each other. Each ring is also coupled to its own bus waveguide, which allows either transmission measurements or pump injection.  Figure \ref{fig:device}(b) shows the linear transmission curves of the two rings measured at room temperature by scanning the frequency of a low-power laser. The cross-section of the waveguide from which the microrings are formed is $1.3\ \mathrm{\mu m} \times 600\ \mathrm{nm}$. Unlike most on-chip microresonators that have been used for comb generation, this waveguide dimension supports only single transverse mode. The radii of the two rings are different (main: $\mathrm{60\ \mu m}$; auxiliary: $\mathrm{58\ \mu m}$), corresponding to different free spectral ranges (FSRs) (main: 378 GHz; auxiliary: 391 GHz). Figure \ref{fig:device}(c) shows the FSR of the main ring versus the relative mode number measured by using frequency comb assisted spectroscopy \cite{r33}. The dispersion is determined from the slope of the curve and is given by $D_2/\left(2\pi\right) \approx -16\ \mathrm{MHz}$, where the $D_2$ parameter signifies the change in FSR from one pair of resonances to the next \cite{r14,r17}. This corresponds to $\beta_2 \approx 130\ \mathrm{ps^2/km}$ and is clearly in the normal dispersion regime.

It can be observed in Fig. \ref{fig:device}(c) that the approximately linear variation of the FSR curve is disturbed around mode $-6$. This is a signature of coupling between the modes of the two rings when their resonant frequencies are close to each other \cite{r29,r34,r33} (see the mode crossing area around 189.5 THz in Fig. \ref{fig:device}(b)). The resonances of the auxiliary ring can be thermally shifted via a microheater. Figures \ref{fig:trans_2D}(a) and \ref{fig:trans_2D}(b) show the zoomed-in transmission spectra versus the heater power for the auxiliary ring and the main ring, measured at their respective bus waveguides. The thermal shifting efficiency of the auxiliary ring is $-600\ \mathrm{MHz/mW}$. The resonances of the main ring are also shifted slightly ($-30\ \mathrm{MHz/mW}$) as no special efforts were made to optimize the thermal isolation.  The resonances in the mode crossing region are split due to the coupling between the rings, a behavior similar to that in quantum mechanical perturbation theory \cite{r35}. The two new resonances evident in the transmission spectra push each other leading to avoided crossings. The effect of mode interaction decreases for heater powers at which the (unperturbed) resonant frequencies of the main ring and auxiliary ring are further separated. Away from the mode-crossing region, one of the resonant features disappears from the transmission spectrum; the remaining resonance recovers to the case with nearly no mode coupling. Naturally, which resonant feature remains outside of the mode-crossing region depends on whether the spectrum is measured from the upper or lower bus waveguide. By changing the heater power, we are able to selectively split and shift an individual resonance.

Denote the field amplitude in the main and auxiliary rings $a_1$ and $a_2$ with time dependences $\exp \left( j\omega_1 t \right)$ and $\exp \left( j\omega_2 t \right)$, respectively. The fields obey the following coupled-mode equations \cite{r36,r37}
\begin{eqnarray}
\frac{\mathrm{d}a_1}{\mathrm{d}t} & = & \left( j\omega_1 - \frac{1}{\tau_1} \right) a_1 + j\kappa_{12} a_2, \label{equ:coupled1}\\
\frac{\mathrm{d}a_2}{\mathrm{d}t} & = & \left( j\omega_2 - \frac{1}{\tau_2} \right) a_2 + j\kappa_{21} a_1, \label{equ:coupled2}
\end{eqnarray}
where $1/\tau_1$, $1/\tau_2$ are the decay rates of the main and auxiliary ring, respectively; $\kappa_{12}=\kappa_{21}^*=\kappa$ is the coupling rate. $\tau_1$ and $\tau_2$ are related to the loaded cavity quality factors ($Q$'s) by $Q_1=\omega_1\tau_1/2$, $Q_2=\omega_2\tau_2/2$.

\begin{figure}[tbp]
\includegraphics[width=7.5cm]{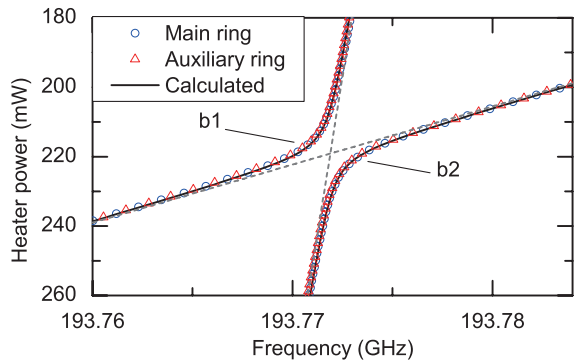}
\caption{\label{fig:reson_shift} Resonant frequencies, measured at the bus waveguide attached to the main ring or to the auxiliary ring, respectively, in the mode-splitting region around 193.77 THz. Branch ``b1'' is red shifted, while branch ``b2'' is blue shifted with respect to the resonant frequencies assuming no mode coupling (the gray dashed lines, derived through linearly fitting the resonant frequencies versus the heater power in the region away from the mode splitting area).}
\end{figure}

The eigenvalues of Eqs. (\ref{equ:coupled1}) and (\ref{equ:coupled2}) are $j\omega^{'}_{1,2}$ where
\begin{eqnarray}
\omega^{'}_{1,2} & = & \frac{\omega_1+\omega_2}{2}+\frac{j}{2}\left( \frac{1}{\tau_1} + \frac{1}{\tau_2} \right) \nonumber \\
        & & \pm \sqrt{\left[ \frac{\omega_1-\omega_2}{2} + \frac{j}{2}\left( \frac{1}{\tau_1} - \frac{1}{\tau_2} \right) \right]^2 + \left|\kappa\right|^2}.
\end{eqnarray}
The real parts of $\omega^{'}_{1,2}$ are the new resonant frequencies affected by mode coupling, and the imaginary parts are the decay rates. Figure \ref{fig:reson_shift} shows the resonant frequencies of the main and auxiliary rings in the mode-splitting region around 193.77 THz. Also shown are the calculated results with $Q_1=7.5\times 10^5$, $Q_2=3.7\times 10^5$, and $\kappa=1.05\times 10^{10}\ \mathrm{s^{-1}}$. Here the $Q$'s are extracted from the linewidths of the transmission resonances measured away from the mode-crossing region, while $\kappa$ is adjusted to fit the data of Fig. \ref{fig:reson_shift}. The agreement between calculation and measurement is excellent. Note that the curves acquired by measuring the transmission at either the top or bottom waveguide agree essentially perfectly, which is expected since in either case we are probing the same hybridized resonances. One of the hybridized resonances is red shifted (branch ``b1") with respect to the original resonances assuming no mode splitting, while the other is blue shifted (branch ``b2''). In comb generation, the resonance shifts play a key role in creating phase matched conditions enabling modulational instability.

For comb generation, the pump laser is slowly tuned into resonant mode 0 of the main ring from the blue side. In this process, the intracavity pump field stays on the upper branch of the bistability curve where modulational instability is absent in a normal-dispersion microresonator undisturbed by mode interactions \cite{r19,r20,r21}. With our coupled-microring scheme,  interactions are intentionally introduced and controlled by thermally tuning the auxiliary ring to facilitate comb generation. At the initial stage of comb formation, essentially only three resonant modes are involved: the pumped mode $\omega_0$ and the two sideband modes $\omega_{\pm \mu}$ where new frequency components first start to grow. The effect of dispersion corresponds to unequal distances between these three modes ($\omega_\mu - \omega_0 \ne \omega_0 - \omega_{-\mu}$). Define the resonance asymmetry factor by $\Delta^2 \omega = \omega_\mu - \omega_0 - \left( \omega_0-\omega_{-\mu} \right)$ which is a generalization of the $D_2$ parameter to a form relevant to modulational instability at the $\pm\mu^{\mathrm{th}}$ resonant modes. Note that positive $\Delta^2 \omega$ corresponds to an anomalous dispersion, for which modulational instability may occur, while negative $\Delta^2 \omega$ corresponds to normal dispersion. By selectively splitting one of the sideband modes $-\mu$ or $\mu$, the asymmetry factor may be changed so that an equivalent anomalous dispersion is achieved ($\Delta^2 \omega$ becomes positive) involving the blue shifted sideband mode (branch ``b2" in Fig. \ref{fig:reson_shift}). Modulational instability thus becomes possible, leading to generation of initial comb lines. More comb lines can be generated through cascaded four-wave mixing resulting in a broadband frequency comb.

Note that the equivalent dispersion achieved with our scheme is different from the general microresonator dispersion which is represented to the first order by an FSR change linear in mode number. As pointed out above, splitting and shifting of just one resonance is enough for the modulational instability to occur; this is distinct from the usual smooth change of FSR over a series of resonant modes. As a result our scheme is easy to implement and offers an alternative to wideband dispersion engineering \cite{r9,r38}. The equivalent dispersion is related to the resonance asymmetry factor by $\beta_{2,\mathrm{eff}} = -n_0 \Delta^2 \omega /\left( 4\pi^2 c \cdot \mu^2\mathrm{FSR}^2_0 \right)$ where $n_0$ is the refractive index, $c$ the speed of light in vacuum, and $\mathrm{FSR_0}$ is the FSR (Hz) at $\omega_0$. For the dual coupled SiN microrings we employed, the resonance shifts due to mode splitting can be several GHz. When $\mu = 1$, this corresponds to an equivalent dispersion change on the order of $10^4\ \mathrm{ps^2/km}$, which is very difficult to achieve by tailoring the microresonator geometry \cite{r9,r38}. The large equivalent dispersion makes it possible to directly generate 1-FSR combs as the peak frequency of modulational instability gain is inverse to the dispersion \cite{r19,r21}. This may prove especially useful for large microresonators with small FSRs \cite{r29}, where without mode interactions achieving direct 1-FSR comb generation is difficult. Direct generation of 1-FSR combs offers the advantage of easy to access phase locked states \cite{r14,r22,r23,r26}.

\begin{figure}[tbp]
\includegraphics[width=8.8cm]{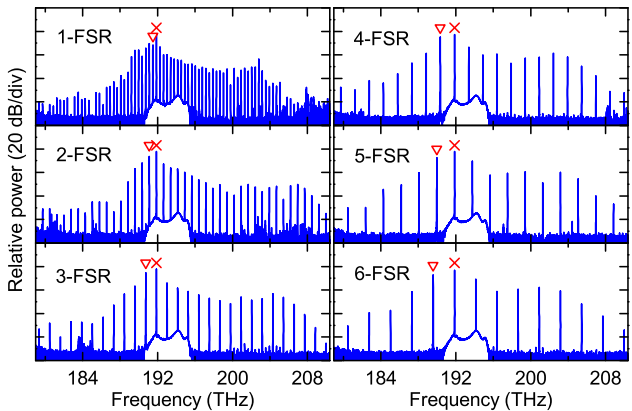}
\caption{\label{fig:tunable_comb} Repetition rate selectable comb. The laser pumps mode 0 of the main ring (marked with $\times$) and the frequency is fixed after tuning into the resonance. The off-chip pump power is 1 W (0.5 W on chip). The auxiliary ring is thermally tuned to facilitate comb generation. The heater power is 108 mW (1-FSR), 95 mW (2-FSR), 68 mW (3-FSR), 44 mW (4-FSR), 20 mW (5-FSR), and 0 mW (6-FSR), respectively. The frequencies at which strong mode interactions occur are marked with $\triangledown$.}
\end{figure}

By controlling the mode interaction location, the initial comb lines can be selectively generated at specified resonances, giving rise to a repetition-rate-selectable comb. Figure \ref{fig:tunable_comb} shows the results when the off-chip pump power is 1 W (0.5 W on chip). The comb repetition rate is tuned from 1-FSR ($\sim378\ \mathrm{GHz}$) to 6-FSR ($\sim2.27\ \mathrm{THz}$) while maintaining a similar comb spectral envelope. All the combs exhibit low intensity noise (below our experimental sensitivity) directly upon generation. High coherence is thus anticipated \cite{r22,r23}.

\begin{figure}[tbp]
\includegraphics[width=8.6cm]{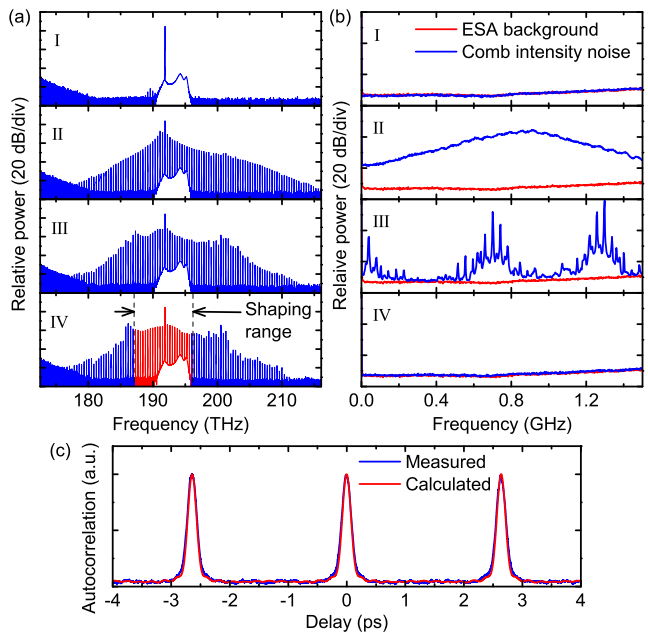}
\caption{\label{fig:mode_locking} Mode-locking transition in the normal dispersion region with dual coupled microrings. The laser pumps mode 0 of the main ring. The off-chip pump power is 1.8 W (0.9 W on chip). (a) Comb spectra when the heater power is 98 mW (\uppercase\expandafter{\romannumeral1}), 109 mW (\uppercase\expandafter{\romannumeral2}), 115 mW (\uppercase\expandafter{\romannumeral3}), and 118 mW (\uppercase\expandafter{\romannumeral4}), respectively. (b) Corresponding comb intensity noise. ESA: electrical spectrum analyzer. (c) Autocorrelation of the transform-limited pulse after line-by-line shaping. The lines in the lightwave C+L band are shaped. The pump line is suppressed by 10 dB to increase the extinction ratio of the compressed pulse.}
\end{figure}

Mode-locking transitions, in which the comb first shows high intensity noise and then transitions to a low-noise state, are also observed. Figure \ref{fig:mode_locking} shows the results with 1.8 W off-chip pump power. The strongest mode interaction is experienced by the first mode to the red side of the pump. By changing the heater power, the spectrum first shows no comb (Fig. \ref{fig:mode_locking}(a) \uppercase\expandafter{\romannumeral1} and Fig. \ref{fig:mode_locking}(b) \uppercase\expandafter{\romannumeral1}), then a 1-FSR comb which at this pump power is generated with high intensity noise (Fig. \ref{fig:mode_locking}(a) \uppercase\expandafter{\romannumeral2}, \uppercase\expandafter{\romannumeral3} and Fig. \ref{fig:mode_locking}(b) \uppercase\expandafter{\romannumeral2}, \uppercase\expandafter{\romannumeral3}). By further optimizing the heater power, the comb intensity noise suddenly drops below the equipment level (Fig. \ref{fig:mode_locking}(a) \uppercase\expandafter{\romannumeral4} and Fig. \ref{fig:mode_locking}(b) \uppercase\expandafter{\romannumeral4}). The low-noise comb is relatively broad with a 10-dB bandwidth of around 15 THz (120 nm). To further verify the mode-locked state of the comb, a fraction of the comb spectrum is selected and phase compensated line-by-line by using a pulse shaper in the lightwave C+L band. The autocorrelation trace of the compressed pulse, shown in Fig. \ref{fig:mode_locking}(c), is in good agreement with the calculated result assuming ideal phase compensation (autocorrelation width 200 fs, corresponding to ~130-fs pulse width). The ability to fully compress to a transform-limited pulse suggests high coherence of the low-noise comb \cite{r22}.

In summary, we have demonstrated a novel scheme incorporating coupled microresonators for Kerr frequency comb generation in the normal dispersion region.  The dynamics of comb formation can be controlled by adjusting the interactions between the microresonators. Repetition rate selection and broadband mode-locking transitions are achieved with coupled SiN microrings. By incorporating a microheater to tune the dual rings into resonance and avoiding accidental degeneracies associated with few-moded waveguides, our scheme shows for the first time a reliable design strategy for normal-dispersion microcombs. As normal dispersion is easy to access in most nonlinear materials, the proposed scheme facilitates microcomb generation and is of particular importance in wavelength ranges (e.g., the visible) where the material dispersion is likely to dominate.

\acknowledgments This work was supported in part by by the Air Force Office of Scientific Research under grant FA9550-12-1-0236 and by the DARPA PULSE program through grant W31P40-13-1-0018 from AMRDEC.

\bibliography{refs}

\end{document}